\documentclass{PoS}

\usepackage{amsmath}

\newcommand\beq{\begin{eqnarray}}
\newcommand\eeq{\end{eqnarray}}
\newcommand\Tab[1]{Table~\ref{tab:#1}}
\newcommand\Fig[1]{Fig.~\ref{fig:#1}}

\title{Progress in the lattice simulations of Sp($2N$) gauge theories}

\ShortTitle{Lattice Sp($2N$)}

\author{\speaker{Jong-Wan Lee},$^{ab}$\thanks{Supported in part by the National Research Foundation of Korea grant funded by the Korea government(MSIT) (NRF-2018R1C1B3001379).} \footnotemark[4]
	     Ed Bennett,$^c$\thanks{Funded by the Supercomputing Wales project, which is part-funded by the European Regional Development Fund (ERDF) via Welsh Government.} 
	     Deog Ki Hong,$^a$\thanks{Supported in part by Korea Research Fellowship program funded by the Ministry of Science, ICT and Future Planning through the National Research Foundation of Korea (2016H1D3A1909283).}
	     C.-J.~David Lin,$^{de}$\thanks{Supported by Taiwanese MoST grant 105-2628-M-009-003-MY4.} 
	     Biagio Lucini,$^f$\thanks{Supported in part by the Royal Society and the Wolfson Foundation.} \footnotemark[7]
	     Maurizio Piai,$^g$\thanks{Supported in part by the STFC Consolidated Grants ST/L000369/1 and ST/P00055X/1.} 
             and Davide Vadacchino$^{h}$\\
             \llap{$^a$}Department of Physics, Pusan National University, 
			Busan 46241, Korea\\
             \llap{$^b$}Extreme Physics Institute, Pusan National University, 
			Busan 46241, Korea\\ 
             \llap{$^c$}Swansea Academy of Advanced Computing, Swansea University, 
		        Singleton Park, Swansea, SA2 8PP, UK\\
             \llap{$^d$}Institute of Physics, National Chiao-Tung University, 
			Hsinchu 30010, Taiwan\\
             \llap{$^e$}Centre for High Energy Physics, Chung-Yuan Christian University, 
			Chung-Li 32032, Taiwan\\
             \llap{$^f$}Department of Mathematics, Computational Foundary, Bay Campus, Swansea University, 
			Swansea, SA1 8EN, UK\\
             \llap{$^g$}Department of Physics, Swansea University, 
			Singleton Park, Swansea, SA2 8PP, UK\\
	     \llap{$^h$}INFN, Sezione di Pisa, 
			Largo Pontecorvo 3, 56127 Pisa, Italy\\
             E-mail:  \email{jwlee823@pusan.ac.kr}, 
		      \email{e.j.bennett@swansea.ac.uk}, 
		      \email{dkhong@pusan.ac.kr}, 
		      \email{dlin@mail.nctu.edu.tw}, 
		      \email{b.lucini@swansea.ac.uk},
		      \email{m.piai@swansea.ac.uk}, 
		      \email{davide.vadacchino@pi.infn.it}}

\abstract{
We report on the status of our programme to simulate Sp($2N$) gauge theories on the lattice. 
Motivated by the potential realization of an SU($4$)/Sp($4$)$\sim$SO($6$)/SO($5$) composite Higgs model 
and the applications to self interacting dark matter, 
we first perform dynamical simulations of Sp($4$) theories with two Dirac flavors in the fundamental representation. 
Preliminary results of the meson spectrum are presented, along with discussion of the lattice systematics. 
We also introduce two-index anti-symmetric Dirac fermions. Such fermions
are relevant in the context of  partial top compositeness, 
provided they carry SU(3) color quantum numbers, and hence we introduce three (Dirac) copies of them. 
We present the quenched meson spectrum 
and explore the phase space of bare lattice parameters. 
For all the numerical simulations we use the standard Wilson lattice gauge and fermion actions.
}

\FullConference{The 36th Annual International Symposium on Lattice Field Theory - LATTICE2018\\
		22-28 July, 2018\\
		Michigan State University, East Lansing, Michigan, USA.}

\begin{document}

\section{Introduction}
Many phenomenological models in physics beyond the Standard Model (BSM), 
either addressing the (little) hierarchy problem in the Standard Model (SM) 
or searching for theories of 
dark matter, or both,
rely on the existence of novel strong dynamics. 
Some of them may be realised at short distances by  Sp(2N) gauge theories with fermionic matter fields. 
In particular, the theory with two Dirac fermions in the fundamental representation 
in which the (enhanced) global symmetry SU($4$) is broken to Sp($4$) in the presence of 
the fermion condensate and/or finite fermion mass 
naturally realizes the SO($6$)/SO($5$) composite Higgs model, 
where the Higgs doublets of the conventional SM are identified by the four out of 
five pseudo Nambu-Goldstone (NG) bosons corresponding to the coset space \cite{Kaplan:1984}. 
If we add fermions in the two-index anti-symmetric representation to this model, 
we might also explain the mass hierarchy of quarks through the mixing with the 
fermionic composite operators composed of fermions from both representations, 
a mechanism referred to as partial compositeness \cite{Kaplan:1991}. 
In the context of dark matter phenomenology, furthermore, Sp($2N$) theory with $N_f$ fundamental Dirac flavors 
($N_f\geq 2$, but sufficiently samll to be in the asymptotically free range) 
has been considered as one of the candidates for strongly coupled gauge theories 
realising the strongly-interacting-massive-particle (SIMP) mechanism \cite{Murayama:2014}. 

Sp($2N$) theories with $N\geq 2$ are new territory for the lattice community. 
While pure gauge theories were investigated in \cite{Holland:2004} a while ago, 
the theories with fermions were only started being studied by the current authors 
very recently. 
In an earlier publication \cite{Bennett:2018} we described the basic lattice set-up in details, 
restricted attention to Sp($4$) and calculated the spectrum of glueballs and mesons constructed 
by fundamental fermions in the quenched approximation. 
In this work we extend our studies of the meson spectrum by replacing the quenched fermions with dynamical ones.
In addition, we perform first explorations of the theory with antisymmetric fermionic matter. 

\section{The model}

Within Sp($2N$), the global symmetry structure of the phenomenologically interesting composite-Higgs models 
can be in principle simulated on the lattice without major technical difficulties: 
it can be realised for the minimal number of hypercolours $N=2$ by a mixed representation 
content of two fundamental and three anti-symmetric flavours of Dirac fermions \cite{Barnard:2014}. 
Since the representations are pseudoreal and real, respectively, 
the global symmetries of the matter fields are enhanced to SU($4$)$\times$SU($6$). 
The field content of the model in terms of two-component spinor fields is summarized in \Tab{fields}. 
The continuum and lattice theories, as well as the low-energy effective theory (EFT)---
at the next-to-the-leading order in the (degenerate) fermion mass, within the context of the hidden-local-symmetry (HLS)
---were extensively discussed in \cite{Bennett:2018}.

In our numerical simulations we use the standard Wilson lattice action for the gauge links and fundamental Dirac fermions. 
Gauge configurations are generated by employing the Heat-Bath (HB) algorithm for the pure gauge model 
and the hybrid Monte Carlo (HMC) algorithm for the model with  dynamical fermions. 
We use a variant of HiRep code \cite{Debbio:2008} appropriately modified 
to implement the main features of Sp($4$) theory \cite{Bennett:2018}. 
In this work we have the anti-symmetric two-index irreducible representation of Sp($2N$): 
we follow the prescriptions for SU($N$) in \cite{Debbio:2008} except subtracting 
the omega-trace term, 
where our convention for the $2N\times 2N$ symplectic matrix is 
$\Omega=\left[\begin{array}{cc}
0&\mathbb{I}_{N\times N}\\
-\mathbb{I}_{N\times N}&0
\end{array}\right]$. 
All of the quantities measured on the lattice are converted to physical ones by utilizing the L\"{u}scher's gradient flow 
technique \cite{Luscher:2010, Luscher:2011}. 
Based on the numerical studies of the scale-setting procedure in \cite{Bennett:2018}, 
we decided to use the derivative of the action density $\mathcal{E}(t)$ at nonzero flow time $t$ \cite{BMW:2012}, 
$\mathcal{W}(t)=t \textrm{d}\mathcal{E}(t)/\textrm{d}t$, as it shows small cut-off dependence than 
the case of $\mathcal{E}(t)$ itself. 
The scale is set by $\mathcal{W}(t)|_{t/a^2=(\omega_0/a)^2}=\mathcal{W}_0$ with an optimal choice of $\mathcal{W}_0=0.35$. 

\begin{table}
\caption{The field contents of the theory. The three columns indicate the
  transformation laws under Sp(4) gauge, global SU(4) and SU(6) symmetries, respectively. 
  $q_F$ and $q_{AS}$ are $2$-component fermions.}
\centering
\label{tab:fields}
\begin{tabular}{|c|c|c|c|c|}
\hline\hline
{\rm ~~~Fields~~~} &$Sp(4)$  &  $SU(4)$ & $SU(6)$\cr
\hline
$V_{\mu}$ & $10$ & $1$ & $1$ \cr
$q_F$ & $4$ & $4$ & $1$ \cr
$q_{AS}$ & $5$ & $1$ & $6$ \cr
\hline\hline
\end{tabular}
\end{table}

\section{Mesons in Sp(4) with $N_f=2$ (dynamical) fundamental Dirac fermions}

\begin{figure}[t]
\begin{center}
\includegraphics[width=.49\textwidth]{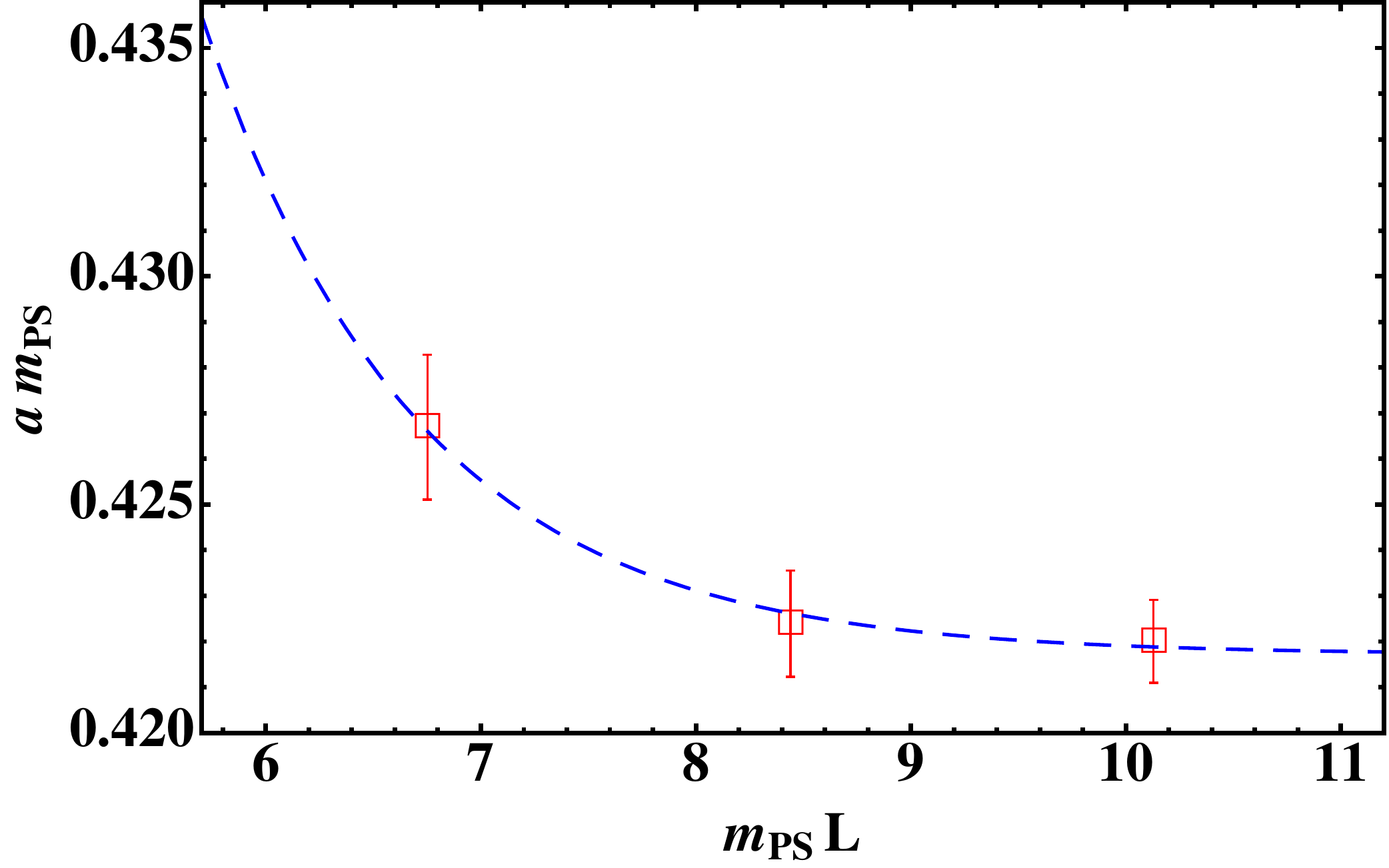}
\includegraphics[width=.49\textwidth]{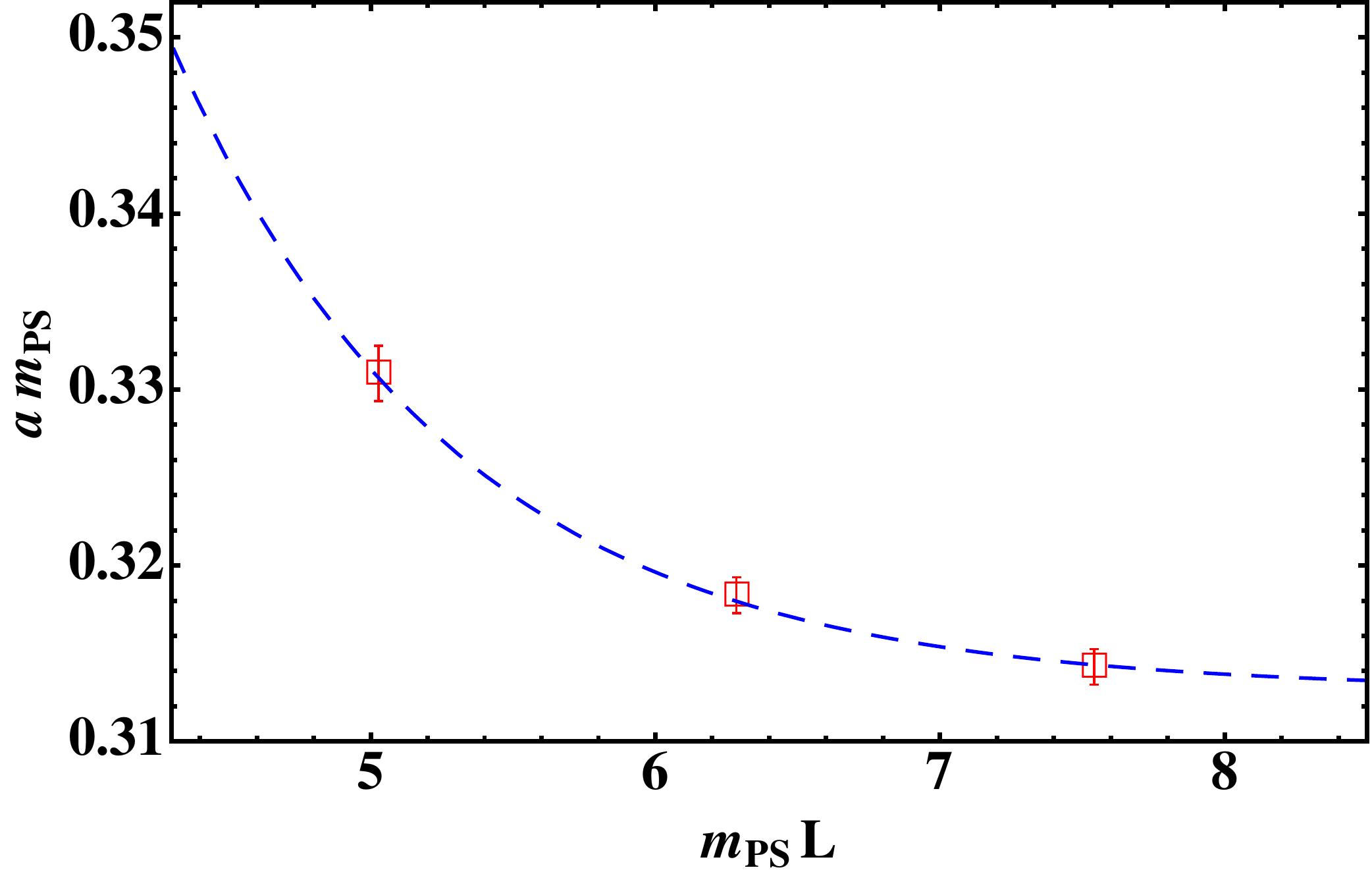}
\caption{%
\label{fig:FV}%
Masses of pseudoscalar mesons measured on various lattice volumes for $m_0=-0.77$ (left) and $-0.79$ (right) 
with $\beta=7.2$. 
The dashed blue lines are the results of exponential fits to the data. 
}
\end{center}
\end{figure}

\begin{figure}[t]
\begin{center}
\includegraphics[width=.49\textwidth]{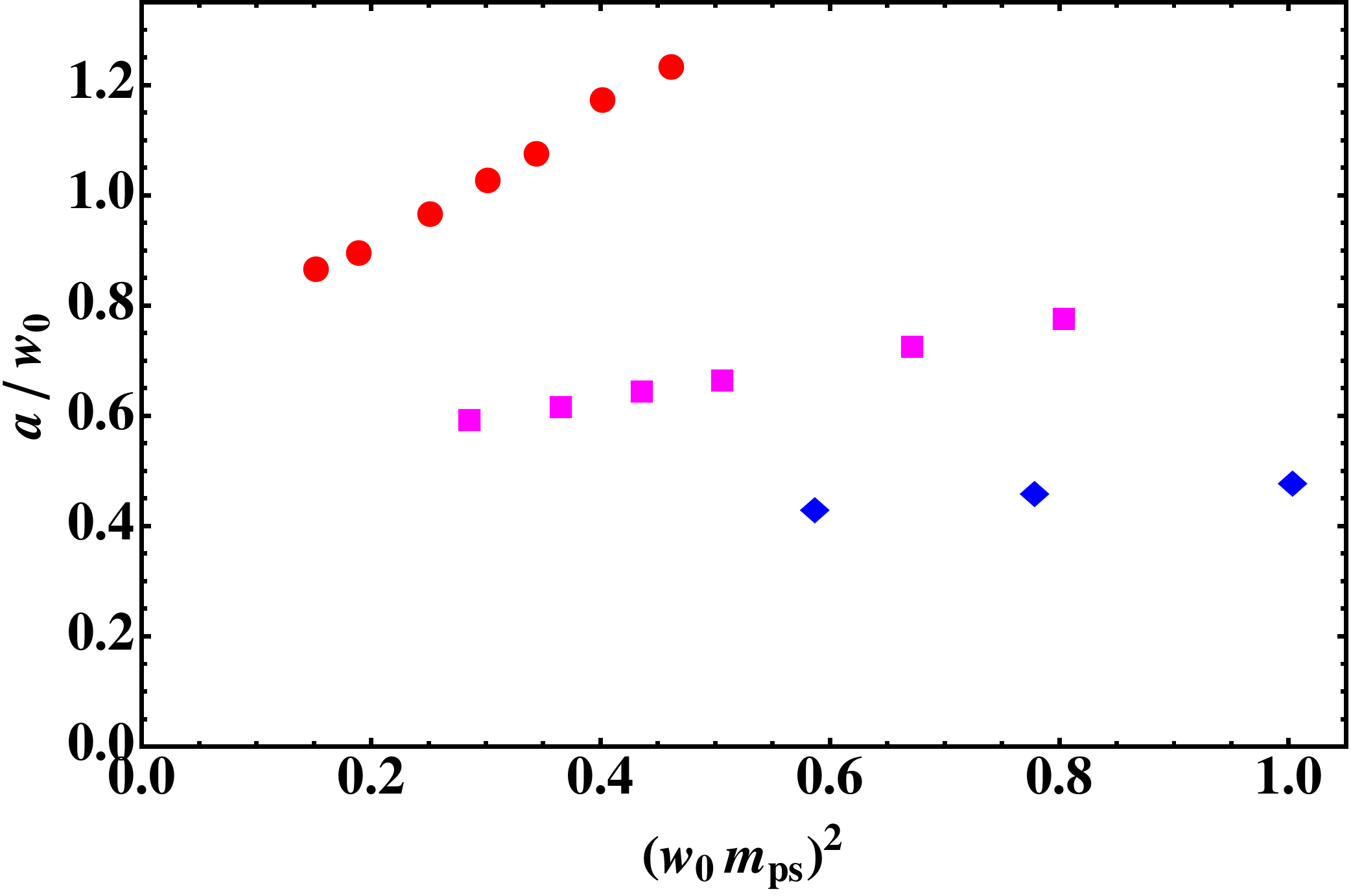}
\caption{%
\label{fig:lattice_spacing}%
Lattice spacings in units of the inverse of the gradient flow scale $w_0$. 
Red circles, purple squares and blue diamonds denote currently available ensembles  
for $\beta=6.9$ with $m_0=[-0.85,-0.924]$, 
$\beta=7.2$ with $m_0=[-0.6,-0.79]$ and $\beta=7.5$ with $m_0=[-0.65,-0.69]$, respectively. 
}
\end{center}
\end{figure}

Our observables of interest are 
the masses and decay constants of flavored pseudoscalar (PS), 
vector (V) and axial-vector (AV) mesons. 
Consider the Euclidean two-point correlation functions 
\beq
C_{\mathcal{O}}(\tau) = \frac{1}{L^3}\sum_{\vec{x}}\langle 0|
\mathcal{O}_M(\vec{x},\tau) \mathcal{O}_M^\dagger (\vec{0},0) |0\rangle 
\xrightarrow{\tau\rightarrow \infty} 
\frac{\langle 0 | \mathcal{O}_M |M \rangle \langle 0 | \mathcal{O}_M |M \rangle^*}{m_M}
\left[e^{-m_M \tau} + e^{-m_M (T-\tau)}
\right].
\eeq
$\mathcal{O}_M(x)=\overline{u}(x) \Gamma d(x)$ 
is the mesonic interpolating operator built with fundamental Dirac fermions $u$ and $d$, 
where the Dirac structures 
result in operators which overlap with PS, V and AV meson states. 
As the matrix elements are proportional to decay constants, 
we parameterize them by
\beq
\langle 0 | \overline{u} \gamma_5 \gamma_\mu d |PS\rangle = f_{PS} p_\mu,~ 
\langle 0 | \overline{u} \gamma_\mu d |V\rangle = f_{V} m_{V} \epsilon_\mu,~ 
\langle 0 | \overline{u} \gamma_5 \gamma_\mu d |AV\rangle = f_{AV} m_{AV}\epsilon_\mu, 
\eeq
where $\epsilon_\mu$ is the unit polarization transverse to the four-momentum $p_\mu$. 
Note that our definitions of the meson states $|M\rangle$ involve the self-adjoint isospin fields 
rather the charged meson fields, where the analogous pion decay constant in QCD is $f_\pi\simeq 93$ MeV. 
As the measured decay constants receive multiplicative renormalization, 
we use the lattice perturbation theory at the one-loop to match the lattice data to the continuum. 
For further details, we refer the reader to Ref. \cite{Bennett:2018}.

Before we present the numerical results of the mass spectrum, 
we discuss the systematic errors due to the lattice artefacts. 
First of all, we investigate the finite volume effects. 
In \Fig{FV}, we show the masses of the PS meson in lattice units for $\beta=7.2$ with $m_0=-0.77$ and $-0.79$ 
on three different lattice volumes, $L/a=16,\,20,\,24$. 
As the Sp(4) theory with $N_f=2$ fundamental Dirac fermions is expected to exhibit color confinement, 
meson masses will receive finite volume corrections in the form of an exponential fall-off 
whose characteristic decay rate will be the mass of the lightest excitation in the theory, i.e. the mass of PS meson. 
We therefore perform a fit of the data to the function, 
$m_{PS}(L)=m_{PS}^\infty +A e^{-m_{PS}^\infty L}$, 
where the fit results are denoted by blue dashed lines. 
With a statistical uncertainty of $\sim 0.3\%$ the size of FV effects become compatible with the statistical errors 
at $m_{PS}\, L \sim 7.5$, and thus we choose the lattice volume for all ensembles to satisfy 
the condition $m_{PS}\, L \gtrsim 7.5$. 

As we discussed above 
we achieve the scale setting 
by using the L\"{u}scher's gradient flow techniques. 
In contrast to the lattice QCD, we have already showed that the gradient flow scale $w_0/a$ rapidly changes 
as we vary the fermion mass \cite{Bennett:2018}, see also \cite{Arthur:2016,TACoS:2018}. 
In \Fig{lattice_spacing} we present the lattice spacings 
with respect to the PS mass-squared in units of $w_0$ 
for currently available ensembles. 
The dependence on the fermion mass becomes milder as we approach the continuum. 
On the other hand, the fermion-mass dependence seems to persist in the chiral limit. 
In this article we do not attempt to take the continuum limit due to the limited number of ensembles, 
but postpone it to our future work 
in which we will fully take advantage of the low-energy EFT developed in \cite{Bennett:2018}. 

\begin{figure}[t]
\begin{center}
\includegraphics[width=.49\textwidth]{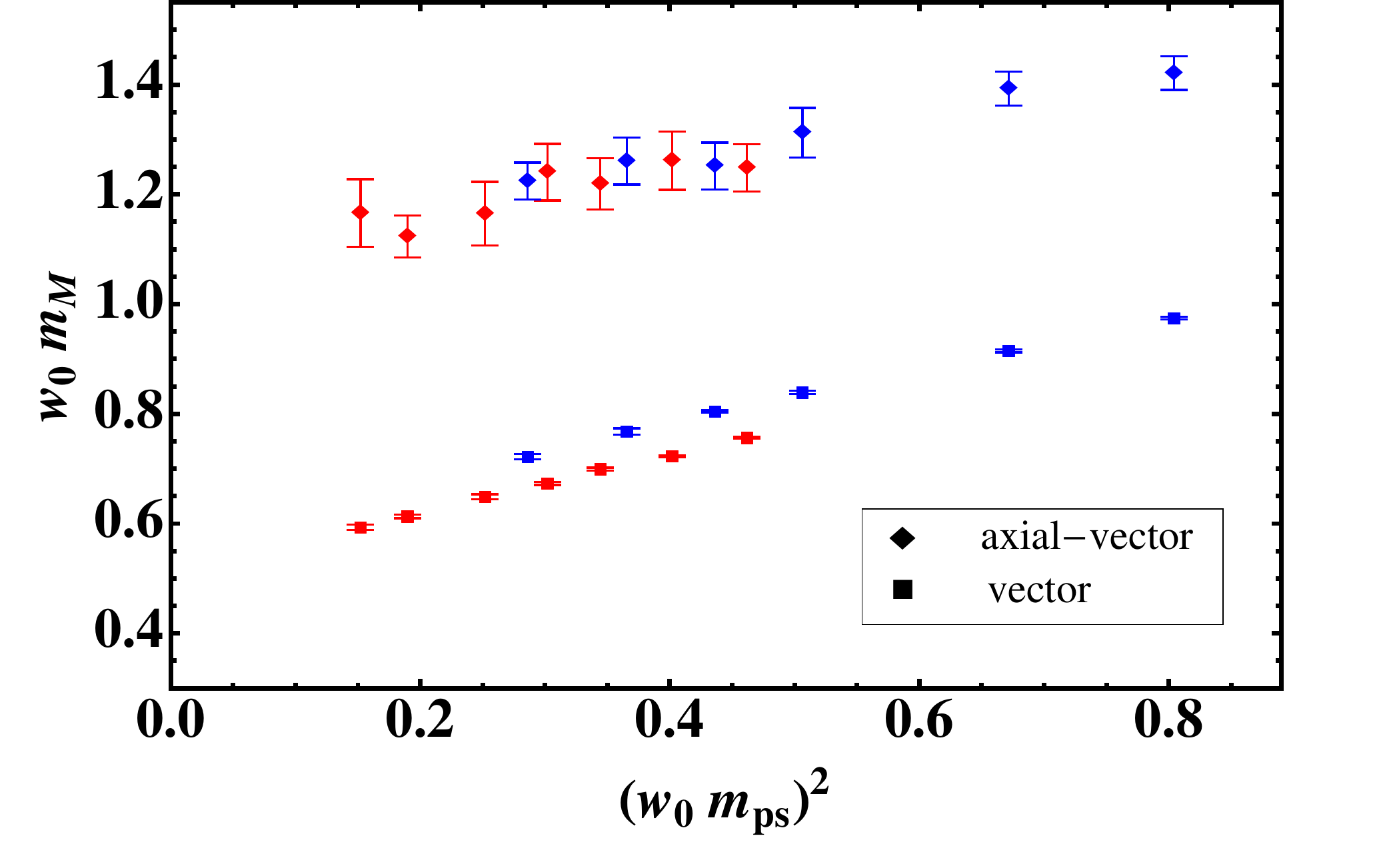}
\includegraphics[width=.49\textwidth]{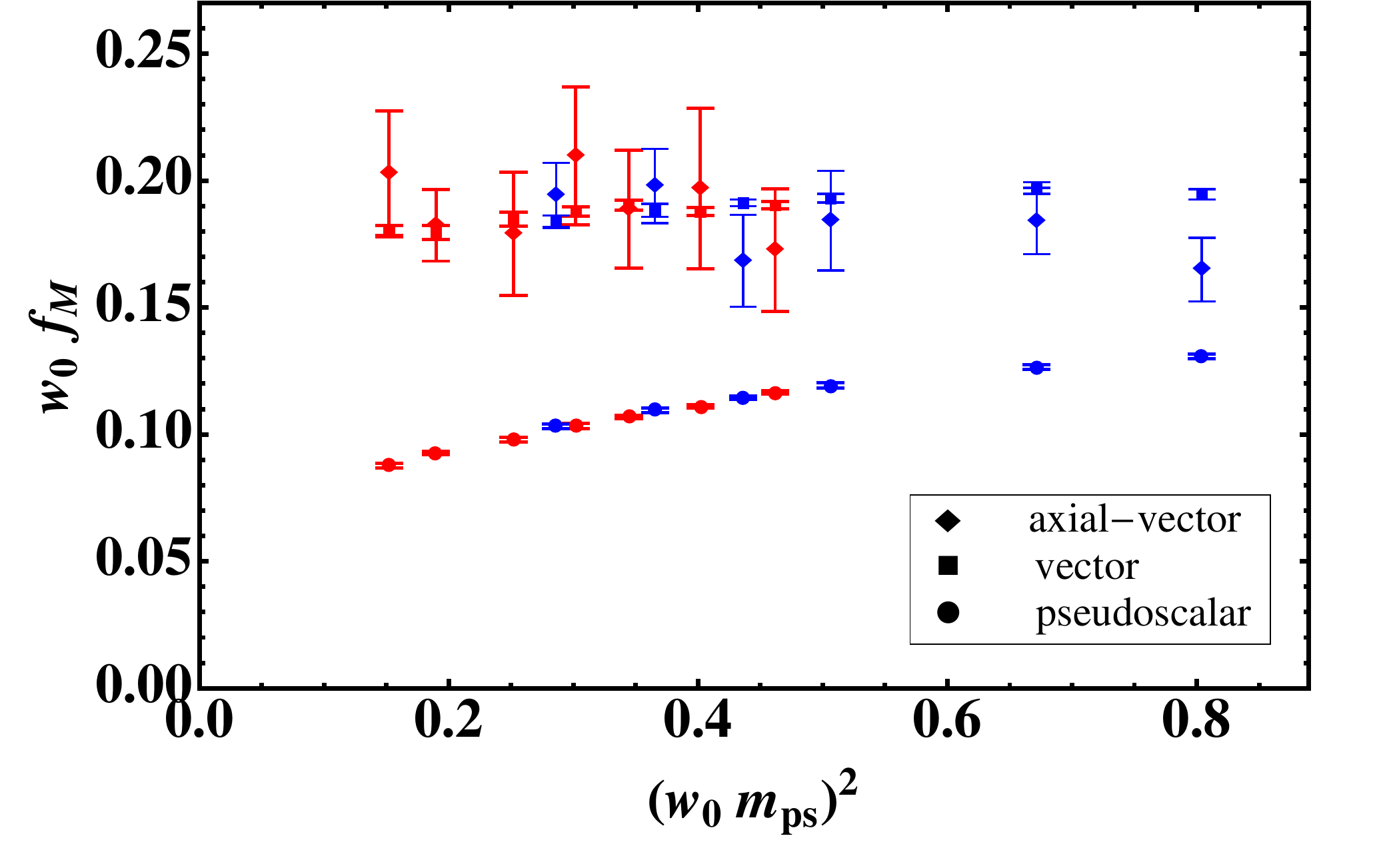}
\caption{%
\label{fig:fund_spectrum}%
Masses and decay constants of PS (circle), V (square) and AV (diamond) mesons 
in the Sp(4) theory with two dynamical Dirac fermions in the fundamental representation. 
We generated ensembles for $\beta=6.9$ on a $32\times 16^3$ lattice, 
except for the two lightest-mass ensembles on a $32\times 24^3$ lattice, 
and for $\beta=7.2$ on a $36\times 16^3$ lattice, 
except for the three lightest-mass ensembles on a $36\times 24^3$ lattice. 
}
\end{center}
\end{figure}

We now present the preliminary results of the dynamical spectrum of the mesons 
for $\beta=6.9$ and $7.2$ with various values of fermion mass in \Fig{fund_spectrum}. 
The masses (left) and the decay constants (right) are shown with respect to the mass squared of the PS meson, 
which is physical and thus scheme-independent. 
From the comparison of two values of the lattice coupling, 
we find that the V meson masses systematically receive sizable corrections due to the discretization. 
In the case of V and AV mesons, we also find that with sufficiently small $m_{PS}$ 
the decay constants are consistent with each other for given statistical errors, 
and insensitive to $m_{PS}$ and $\beta$. 
From the fact that the V meson mass increases for smaller lattice spacing, 
we expect that the $S$ parameter, 
defined in the framework of the low-energy EFT only with the lightest mesons, 
decreases in the continuum limit. 

\begin{figure}[t]
\begin{center}
\includegraphics[width=.49\textwidth]{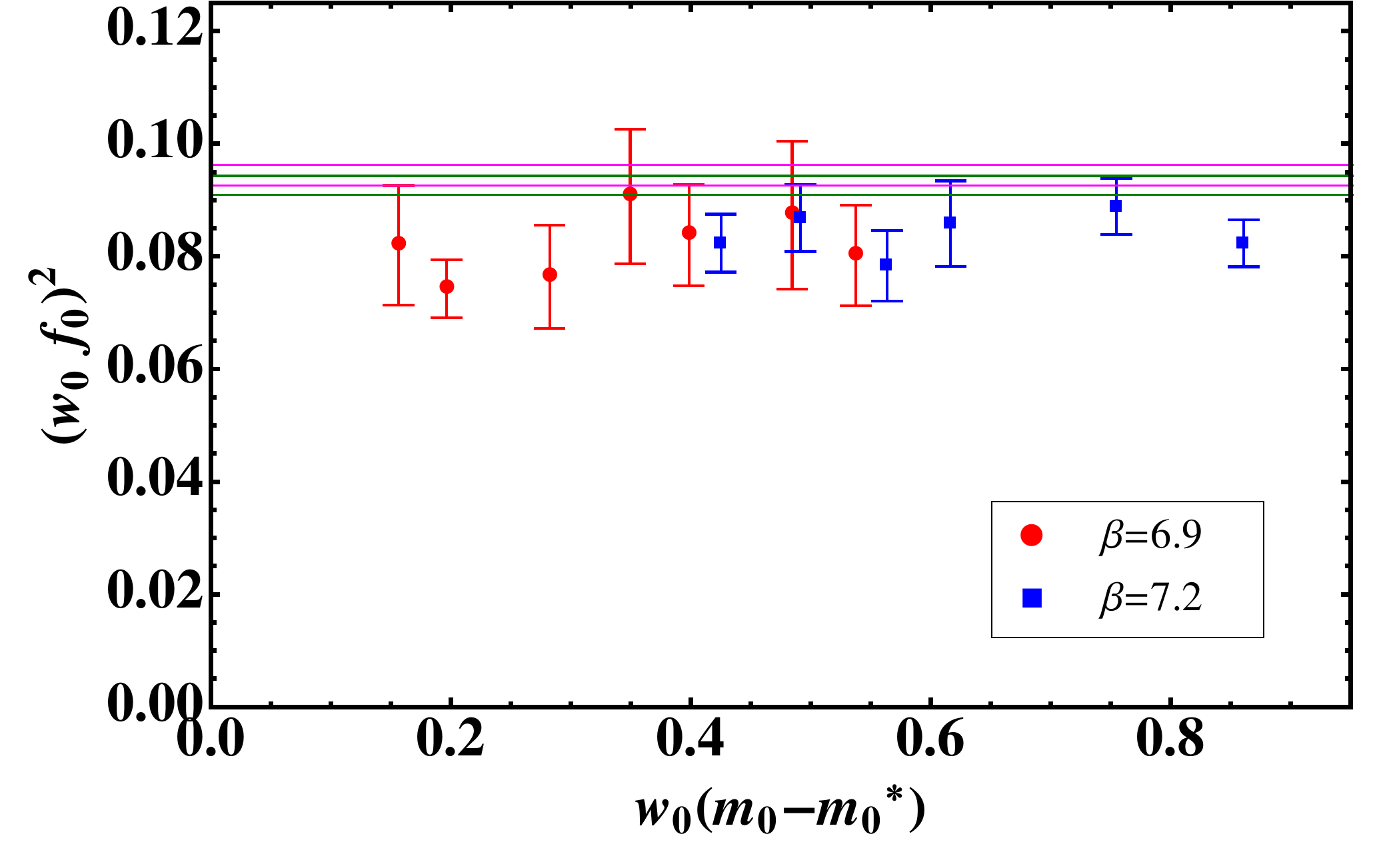}
\caption{%
\label{fig:f0square}%
The Sum of the squared decay constants of the PS, V, AV mesons with dynamical $N_f=2$ fundamental Dirac fermions. 
For reference, we also present the fit results by the colored bands for 
quenched Sp(4) simulations with $\beta=7.62$ (green) and $8.0$ (purple). 
}
\end{center}
\end{figure}

The low-energy NLO EFT predicts that the sum of the squared decay constants of PS, V 
and AV mesons in the zero momentum limit, $f_0^2=f_{PS}^2+f_{V}^2+f_{AV}^2$, 
is independent from the fermion mass, which was evidenced by the quenched spectrum 
\cite{Bennett:2018}. 
In \Fig{f0square} we show the results of $f_0^2$ measured from dynamical simulations 
for various masses, where no significant mass dependence is found over the wide ranges. 
The resulting values are also not statistically far from the results in the quenched cases. 
Such numerical results strongly support predictions of the EFT indicating 
the insensitivity of $f_0^2$ on the fermion mass. 

\section{Mesons in quenched Sp(4) with anti-symmetric Dirac fermions}

\begin{figure}[t]
\begin{center}
\includegraphics[width=.49\textwidth]{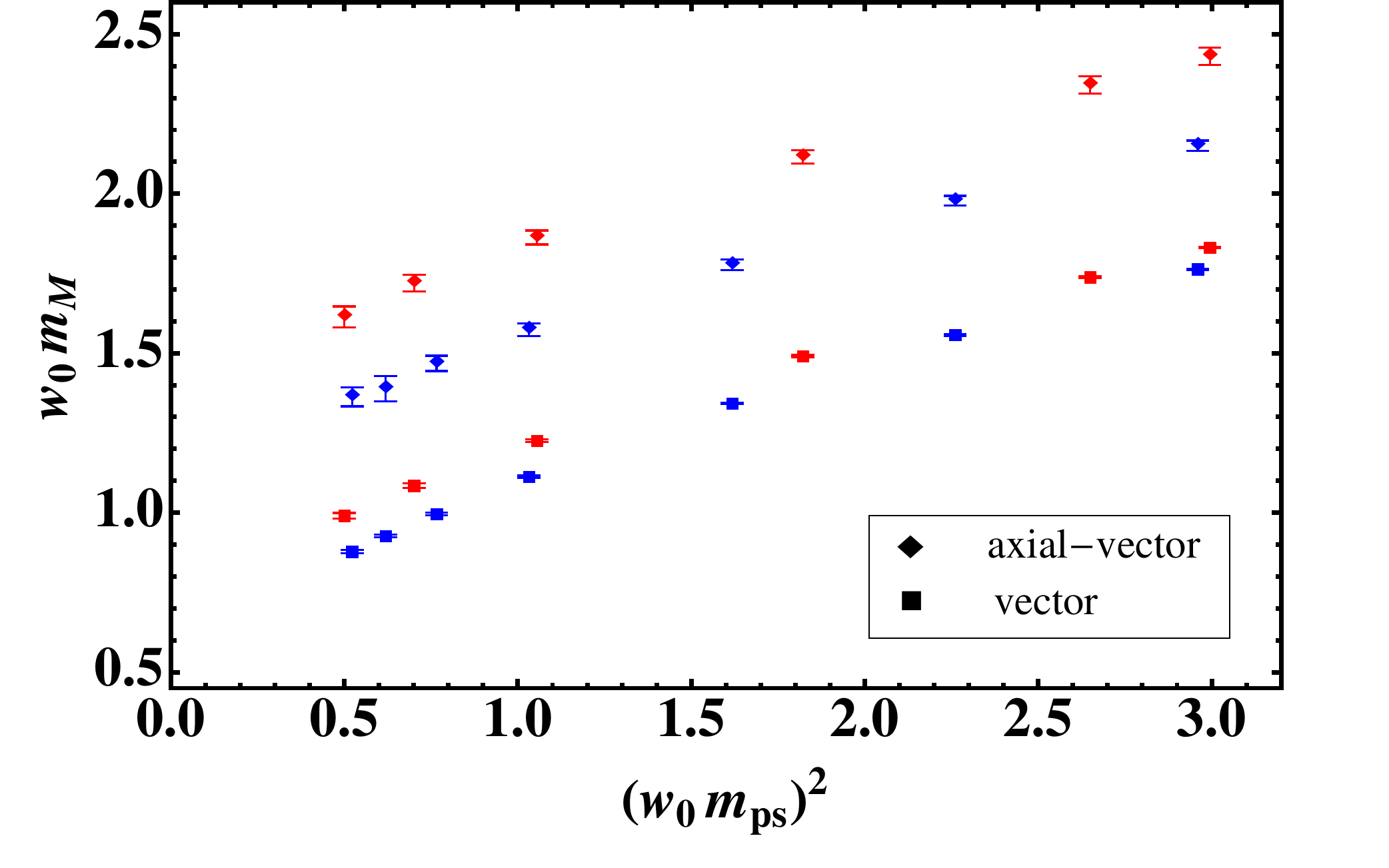}
\includegraphics[width=.49\textwidth]{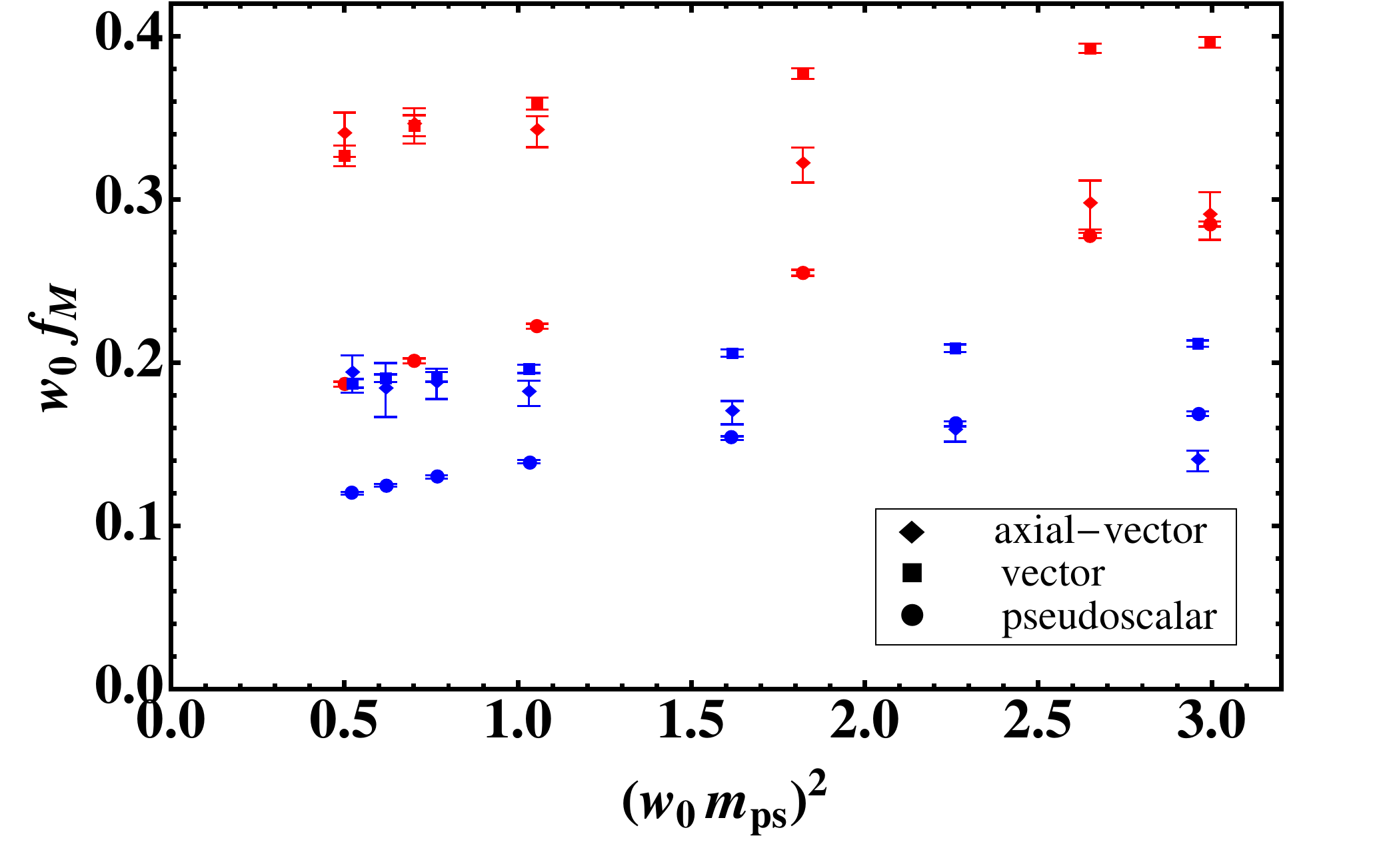}
\caption{%
\label{fig:as_spectrum}%
Masses and decay constants of pseudoscalar (circle), vector (square) and axial-vector (diamond) mesons 
constructed from fermions in the anti-symmetric (red) 
and fundamental (blue) representation in the quenched Sp(4) theory at $\beta=8.0$. 
The lattice volume was $48\times 24^3$. 
}
\end{center}
\end{figure}

\begin{figure}[t]
\begin{center}
\includegraphics[width=.32\textwidth]{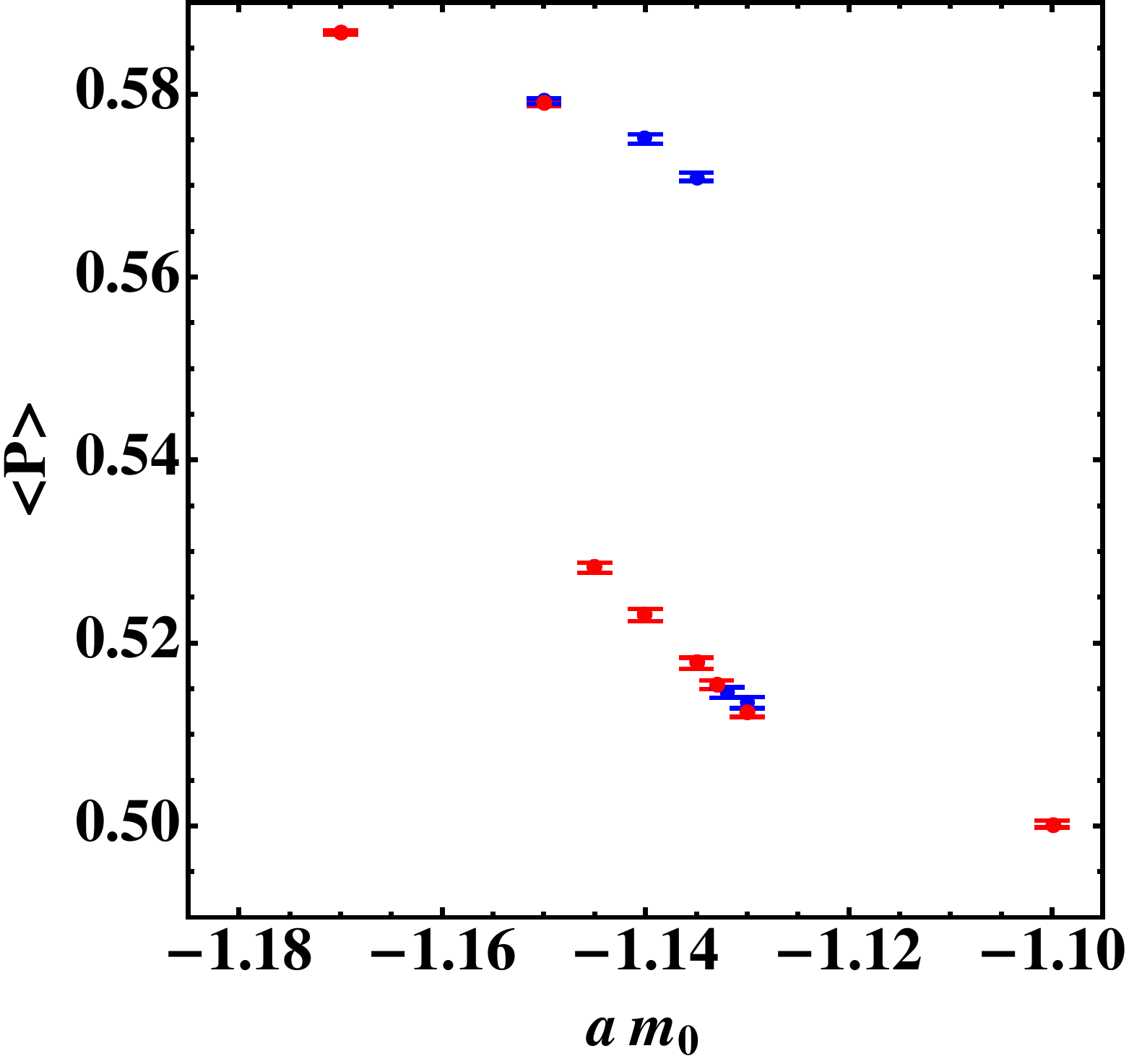}
\includegraphics[width=.32\textwidth]{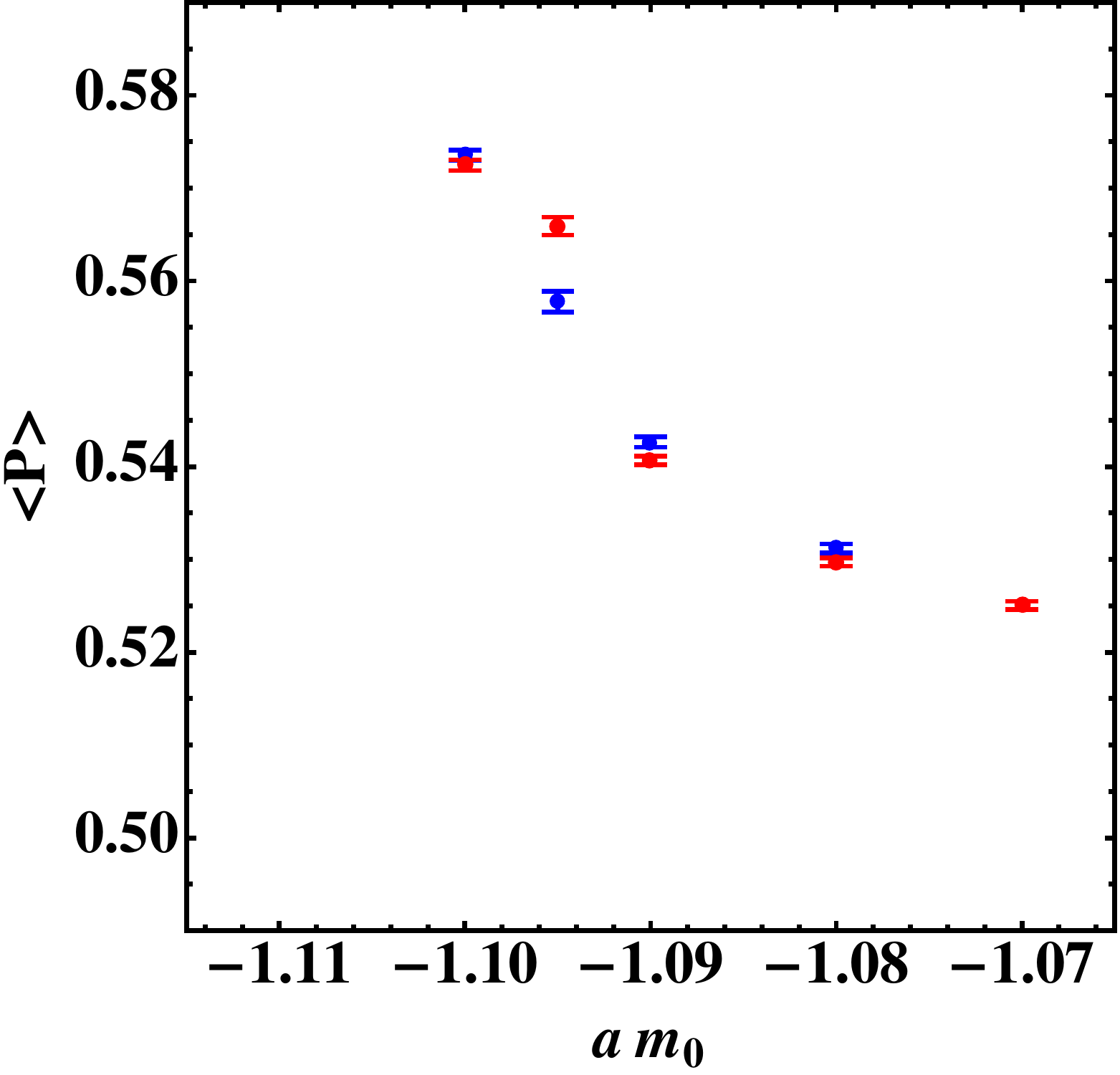}
\includegraphics[width=.32\textwidth]{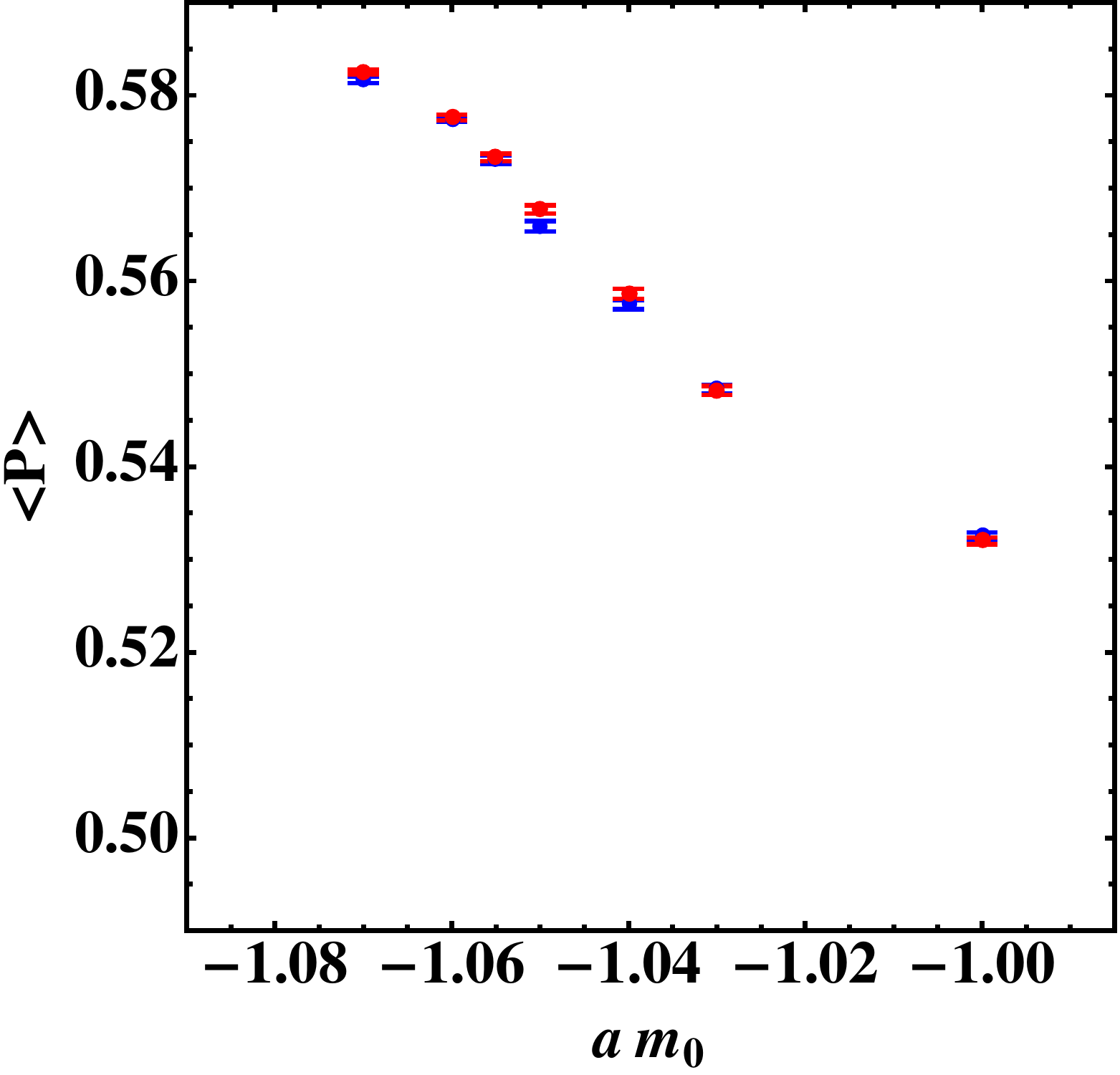}
\caption{%
\label{fig:as_bulk_phase}%
Mass scan of the Sp(4) theory with $N_f=3$ anti-symmetric Wilson fermions 
at $\beta=6.4$, $6.5$ and $6.6$ from left to right, respectively. 
The red and blue symbols denote the expectation values of the plaquette $\langle P \rangle$ obtained from 
random (hot) and unit (cold) initial configurations on a $8^4$ lattice. 
}
\end{center}
\end{figure}

In Sp($2N$) theory the anti-symmetric two-index representation is real, 
hence with $N_f$ Dirac flavours the global symmetry is enhanced to SU($2N_f$) and broken to SO($2N_f$) 
in the presence of non-zero (symmetric) condensate. 
Pseudo NG bosons correspond to $2N_f^2+N_f-1$ broken generators 
which belongs to the coset SU($2N_f$)/SO($2N_f$). 
In terms of Dirac flavors, the NG bosons are 
$N_f^2-1$ mesons in the adjoint representation, and $N_f(N_f+1)/2$ diquarks and anti-diquarks 
in the symmetric representation. 
As in the case with fundamental fermions, we focus on the spectrum of flavored PS, V, 
and AV mesons which are degenerate with the corresponding diquark and anti-diquark states transformed in 
the same way under the global symmetry in the massless limit. 
As dynamical ensembles are not available yet, we calculate the masses and decay constants 
in the quenched limit from the same ensembles used for the fundamental fermions in \cite{Bennett:2018}. 
The results for $\beta=8.0$ are shown as red symbols in \Fig{as_spectrum}. 
For a comparison we also present the results for the quenched spectrum with fundamental fermions 
denoted by blue symbols. 
The masses and decay constants for both representations show similar dependence 
on the PS meson mass, but the overall scale is substantially different. 

Toward the dynamical simulation with anti-symmetric fermions, the primary task is to 
search for any singularity associated with the bulk-phase transition by exploring the bare lattice parameter space. 
Using $8^4$ lattices with $\beta=6.4,\,6.5,\,6.6$, we calculate the expectation values of the plaquette 
$\langle P \rangle$
by varying the bare fermion masses. 
We focus on the vicinity of the region in which the plaquette values change abruptly. 
The results are presented in \Fig{as_bulk_phase}, 
where red squares and blue circles are obtained from the random (hot) and unit (cold) initial configurations. 
We find that hot and cold results are consistent with each other at $\beta=6.6$, 
while they are well separated over the range of $m_0=[-1.145,-1.135]$ at $\beta=6.4$. 
Such strong hysteresis at small $\beta$ provides strong evidence for the existence of a first-order bulk phase transition. 
We therefore estimate our conservative value of the phase boundary as $\beta\gtrsim 6.6$ 
at which the continuum extrapolation can be taken correctly. 

\section{Conclusion}

We presented preliminary results of a first lattice calculation of the meson spectrum 
for the Sp($4$) gauge theory with two dynamical fundamental Dirac fermions. 
We first investigated the systematic effects associated the finite volume and the finite lattice spacing. 
Although the extrapolation to the continuum limit has not been carried out yet, 
our numerical results show consistency with the low-energy EFT expectations. 
We also presented the spectrum for the theory with Dirac fermions in the anti-symmetric representation 
in the quenched limit. Toward the dynamical simulation, 
we explored the phase structure in the lattice parameter space and 
identified the phase boundary for the first order bulk phase transition. 

\section*{Acknowledgements}
We are very grateful to Michele Mesiti and Jarno Rantaharju who assisted us in modifying and improving the HiRep code 
for this project.

\end{document}